\documentclass[conference]{IEEEtran}
\IEEEoverridecommandlockouts
\usepackage{tcolorbox}
\usepackage{fvextra}
\usepackage{cite}
\usepackage{amsmath,amssymb,amsfonts}
\usepackage{algorithmic}
\usepackage{graphicx,dblfloatfix}
\usepackage{textcomp}
\usepackage[x11names, rgb]{xcolor}
\usepackage{orcidlink}

\def\BibTeX{{\rm B\kern-.05em{\sc i\kern-.025em b}\kern-.08em
    T\kern-.1667em\lower.7ex\hbox{E}\kern-.125emX}}

\definecolor{DarkGray}{rgb}{0.282, 0.282, 0.282}
\definecolor{LightGray}{rgb}{0.922, 0.922, 0.922}
\definecolor{Blue}{rgb}{0.098, 0.463, 0.824}

\begin{document}

\title{TACLA: An LLM-Based Multi-Agent Tool for Transactional Analysis Training in Education
\thanks{The work reported in this paper was partly supported by the Polish National Science Centre under grant 2024/06/Y/HS1/00197.}
}

\author{
\IEEEauthorblockN{Monika Zamojska\,\orcidlink{0009-0000-3138-5562}, Jarosław A. Chudziak\,\orcidlink{0000-0003-4534-8652}}
\IEEEauthorblockA{Faculty of Electronics and Information Technology \\
Warsaw University of Technology \\
Warsaw, Poland \\
\{monika.zamojska.stud, jaroslaw.chudziak\}@pw.edu.pl} \\ 
}

\maketitle

\begin{abstract}
Simulating nuanced human social dynamics with Large Language Models (LLMs) remains a significant challenge, particularly in achieving psychological depth and consistent persona behavior crucial for high-fidelity training tools. This paper introduces TACLA (Transactional Analysis Contextual LLM-based Agents), a novel Multi-Agent architecture designed to overcome these limitations. TACLA integrates core principles of Transactional Analysis (TA) by modeling agents as an orchestrated system of distinct Parent, Adult, and Child ego states, each with its own pattern memory. An Orchestrator Agent prioritizes ego state activation based on contextual triggers and an agent's life script, ensuring psychologically authentic responses. Validated in an educational scenario, TACLA demonstrates realistic ego state shifts in Student Agents, effectively modeling conflict de-escalation and escalation based on different teacher intervention strategies. Evaluation shows high conversational credibility and confirms TACLA's capacity to create dynamic, psychologically-grounded social simulations, advancing the development of effective AI tools for education and beyond.
\end{abstract}


\section{Introduction}
The ability to simulate complex human social dynamics represents an important frontier in artificial intelligence \cite{Gao}. Large Language Models (LLMs) combined with Multi-Agent Systems (MAS) have advanced this field, demonstrating unique capabilities in generating human-like conversation \cite{Wang2024}. This progress opens exciting avenues for developing simulation-based tools \cite{Gao2023}. Such tools hold great promise in many sectors, particularly where understanding and practicing difficult social interactions are crucial.

However, despite these advances, current LLM-based agents often lack deep psychological understanding and consistent behavior \cite{Niu2024, Park2023}. While they can copy conversation, they typically struggle to represent distinct, stable personas. They also miss the hidden emotions or complex reasons behind human actions. This limits their ability to create truly challenging and useful environments for social research or training. To move beyond superficial imitation towards more psychologically faithful behavior, agent architectures require the integration of established psychological frameworks that can explain deeper aspects of human interaction.

Transactional Analysis (TA), a psychological theory developed by Eric Berne, offers a compelling framework for addressing these limitations \cite{Berne1972}. TA provides a structured model for understanding personality and interpersonal communication. Its core concept include three distinct ego states — Parent, Adult, and Child — each with unique patterns of thinking, feeling, and behaving \cite{Stewart12}. Furthermore, TA explains how interactions (transactions) unfold between these ego states and form recurring patterns, called psychological \emph{games}. Its well-defined concepts make TA particularly suitable for computational modeling, providing a solid foundation for designing agents capable of complex and psychologically realistic interactions.

The importance of well-developed social and emotional skills is especially paramount for teachers. Educators significantly impact students' confidence and well-being, guiding them in communication, relationship-building, and conflict resolution \cite{Indellicato}. Yet, many teachers report insufficient opportunities for training in these vital interpersonal skills \cite{Shapira}. While Artificial Intelligence (AI) can automate certain routine educational tasks \cite{zhai, WANG, Aljemely}, it cannot replace the uniquely human role of a teacher in providing emotional support or mediating student conflicts \cite{Allam, Guan}. Notably, Transactional Analysis itself has already proven beneficial in educational contexts, helping teachers understand student behavior, improve their classroom interactions, and foster more supportive learning environments \cite{newellbook, Pierzchala, barrow, Stuart}.

To leverage these insights for enhanced teacher training, this paper introduces platform for teacher training, which implements the \textbf{TACLA} (Transactional Analysis Contextual LLM-based Agents) architecture. \textbf{TACLA} is a novel cognitive architecture designed to create agents with psychological depth. This approach integrates core TA principles into an LLM-based Multi-Agent System. Each \textbf{TACLA} agent is modeled as a combination of Parent, Adult, and Child Ego State Agents. These have separate Contextual Pattern Memory and TA-informed reasoning capabilities. This design allows for the creation of realistic Student Agents for classroom scenarios, where teachers can practice communication and management skills. This paper describes the \textbf{TACLA} architecture and its theoretical foundation. It also presents and validates its application as a tool for teacher professional development, though its principles hold potential for broader applications in social simulation.

\section{Background and Related Work}
This section outlines the key theoretical and technological foundations of this research. It begins by overview of the state-of-the-art in LLM-based Multi-Agent Systems and agent architectures, identifying their potential and limitations for social simulation. The text then introduces Transactional Analysis (TA) as the central psychological theory inspiring the approach to agent design in this work. Finally, it illustrates the challenges of classroom dynamics and TA's well-established role in education to provide the practical context for the study.

\subsection{Large Language Models (LLM) based Multi-Agent Systems (MAS)}
Large Language Models (LLMs) are AI systems design to process and generate human-like text. They are transforming education, creating personalized learning materials \cite{Huang}, lesson plans, and test questions \cite{Lamsiyah,Li}. Beyond content creation, LLMs can provide personalized feedback on student work \cite{zhai}, generate reflection triggers \cite{Naik}, and even analyze lessons to provide tips for teachers \cite{Hou}.

When LLMs power AI agents, they enable them to reason, make choices, and interact with others and the environment \cite{tran}. A Multi-Agent System (MAS) combines multiple such agents, each with unique roles \cite{synergymas}. They can cooperate, working on completing a specific task \cite{cinkuszUSA,wawer}. Recent research has focused on examining the applications of MAS in simulating human behavior and social interactions \cite{zhang24,aciids}. The results showed that agents are capable of acting human-like, including organizing social hierarchies \cite{huang25}. However, they have certain limitations, such as reinforcing misuse, social bias, and stereotypes \cite{thapa2025}.

\subsection{Agent Architectures for Social Simulation}
Achieving human-like behavior in social simulations requires agent architectures capable of mimic complex cognitive, emotional, and social processes \cite{Gao}. Traditional agent-based modeling methodologies often rely on rule-based systems or statistical models to define agent behavior \cite{bail2024,flache2022}. With the development of Large Language Models (LLMs), these architecture have seen a significant transformation \cite{gurcan2024}. Modern systems incorporate advanced modules to imitate human intelligence. One of the critical components is memory system \cite{zhong2024,chen2024,huang2024}, which typically involves both short-term memory (within the model’s context window) and long-term memory (persistent information storage often implemented using vector databases or graphs). Retrieving past memories enabling agents to reflect on them, draw conclusions, and use it to guide future behavior \cite{piao2025}. Frameworks like ReAct \cite{yao2023} emphasize planning and reasoning by allowing LLMs to interleave thought generation with action execution. This approach allows for updating action plans based on gathered knowledge and observations.

However, while these architectures enhance agents' capabilities, many still primarily focus on task completion or mapping the reasoning of the human brain \cite{Park2023}. They often lack predefined, deep psychological structures. Studies highlight that LLM-generated responses can exhibit high homogeneity of opinion \cite{ziems2024}. Moreover, agents may be inconsistent in simulated roles, acting in ways that contradict their self-reported traits \cite{huang2024}. In Multi-Agent Systems, agents can adapt to each other linguistically, where one agent's personality may adapt more readily to its counterpart's \cite{frisch}. This suggests the need for architectures that allow for psychological consistency and a deeper exploration of human-like social interactions.

\subsection{Transactional Analysis}
\begin{figure}[b]
\centering
\includegraphics[width=1\columnwidth]{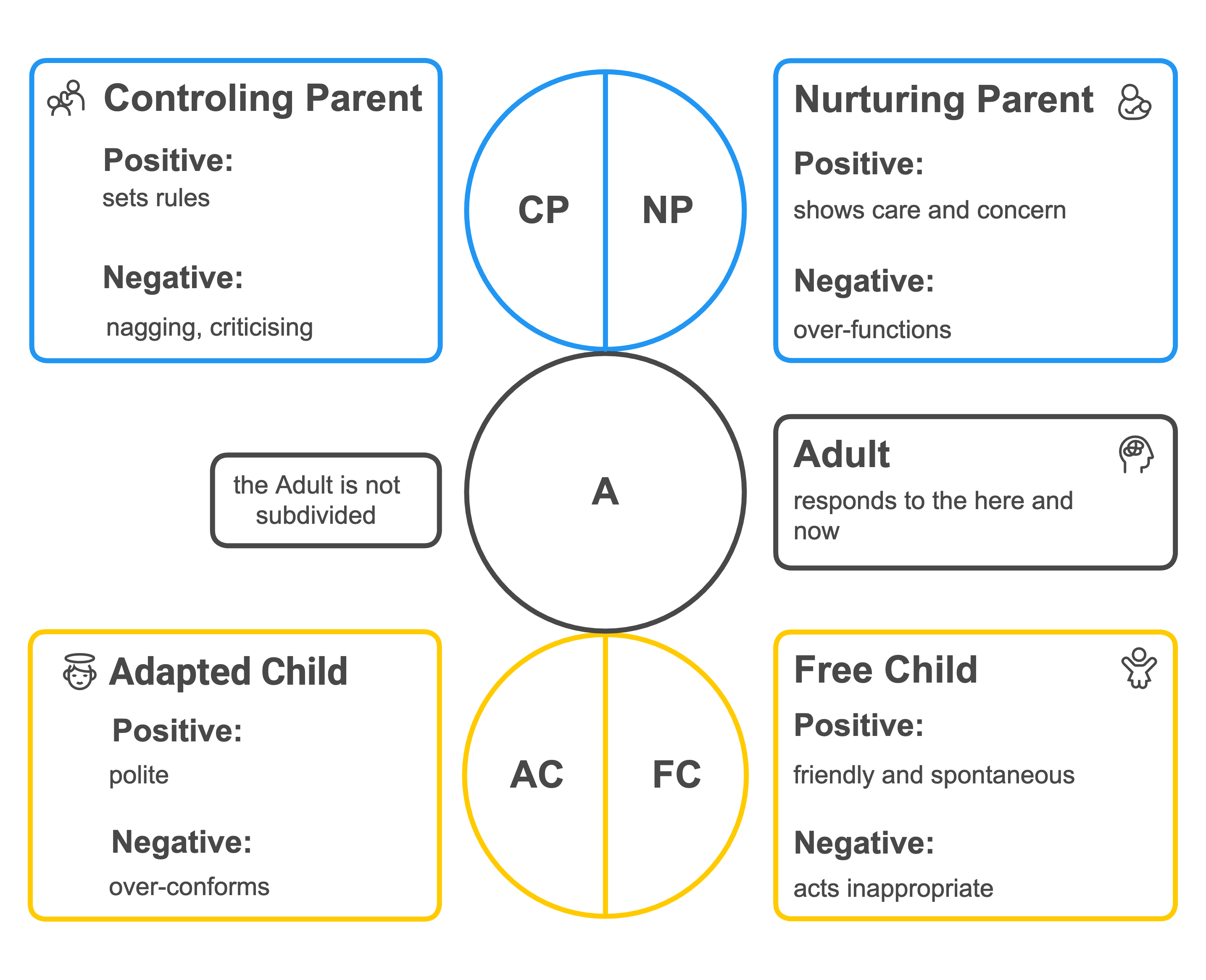}
\caption{The functional model of Ego States with descriptions, based on \cite{newellbook}.}
\label{fig:functional_model}
\end{figure}

Developed by Eric Berne \cite{Berne58}, Transactional Analysis (TA) can provide useful resources to understand and improve interactions. Its structural model identifies three ego states: Parent, Adult, and Child. The Parent state reflects behaviors, thoughts, and emotions adopted from parental figures. The Adult state is grounded in present reality, responding logically to the current situation. The Child state consists of behaviors, emotions, and thought patterns from early childhood experiences \cite{Stewart12}. These states can be further divided into functional subcategories, as shown in Fig. \ref{fig:functional_model}. 

Effective communication occurs when \emph{transactions} (interactions) remain complementary, meaning the response aligns with the expected ego state. When a response comes from a different state, a crossed transaction happens, breaking the flow of communication. For example, if a student operates from the Rebellious Child state (a part of Adapted Child), reacting with resistance or aggression, a teacher may instinctively respond from the Controlling Parent state. This often makes the student continue their behavior and remain in the same state. A more effective approach would be responding from the Adult state, using logic and calm reasoning to induce the state change \cite{pankowska}.

TA is built on several fundamental assumptions. It holds that people are fundamentally \emph{"OK"}, everyone has the capacity to think, and each person can decide their own destiny. Besides, TA introduces key concepts such as \emph{strokes} — small units of recognition that satisfy the need to be noticed — and \emph{games}, repetitive patterns of interaction driven by hidden motivations and predictable outcomes \cite{Berne64}. These ideas help explain how students form beliefs and react in different situations. Teachers can use TA in the classroom to better understand the emotional factors influencing behavior and to separate a student's identity from their actions \cite{newellbook}.

\subsection{Classroom Dynamics: Challenges and the Role of TA}
Teachers face many challenges in managing classrooms. Sometimes, students do not follow the rules, and their disruptive influence affects the whole class \cite{Lloyd}. Dealing with such behavior can cause teacher frustration. It is overwhelming and stressful, especially in overcrowded classrooms \cite{Vakili}. In these situations, it is more difficult to focus on individual students, motivate them, and address their needs. These challenges require teachers to be adaptable and skilled in creating a positive and stable classroom environment.

Teacher training is essential to overcome these challenges \cite{Mustafa}. Research shows that practicing classroom management strategies helps teachers build higher-quality relationships with students \cite{Kumari,Bergin}. When educators receive training in social-emotional skills, they learn how to navigate conflicts, reduce aggression, and create a cooperative, supporting environment \cite{Caires,Shapira}. This training also helps reduce negative feelings such as stress and burnout \cite{Scheirlinckx}. Improved teacher training leads to better student engagement and academic outcomes, while also supporting the development of students' social and emotional skills \cite{chow}.

According to research, teachers who apply TA gain a better understanding of their students and build stronger relationships with them \cite{Stuart}. TA-informed educators are more effective in resolving conflicts and preventing violence \cite{pankowska}. They learn to choose responses that support both emotional and academic development \cite{Alipieva}. By understanding the psychological roots of behavior, teachers can cultivate mutual respect and understanding in the classroom \cite{barrow}. Ultimately, TA empowers both teachers and students with essential life skills \cite{Connors}.

\section{Approach: Modeling Social Dynamics with Transactional Analysis}
To create agents capable of authentic social interaction, this work proposes \textbf{TACLA} (Transactional Analysis Contextual LLM-based Agents), a model that integrates a psychological theory with a structured cognitive framework. This section outlines the conceptual foundations of this approach.

\subsection {Transactional Analysis as a Framework}
Transactional Analysis (TA) provides a structured and powerful framework for analyzing personality and interpersonal communication \cite{Stewart12}. It helps break down complex human behavior into definable components. \textbf{TACLA} specifically adapts TA's concepts to construct agent persona. The core is the model of three distinct ego states — Parent, Adult, and Child. Each ego state represents a distinct way of thinking, feeling, and behavior, directly informing how an agent will process information and generate responses. Beyond ego states, TA introduces other important ideas for agent design. These include \emph{drivers}, which are internal messages that affect behavior, and \emph{life scripts} - unconscious plans formed in early childhood that direct future behavior.

\subsection {The TACLA Conceptual Model}
\begin{figure}[b]
  \centering
  \includegraphics[width=\columnwidth]{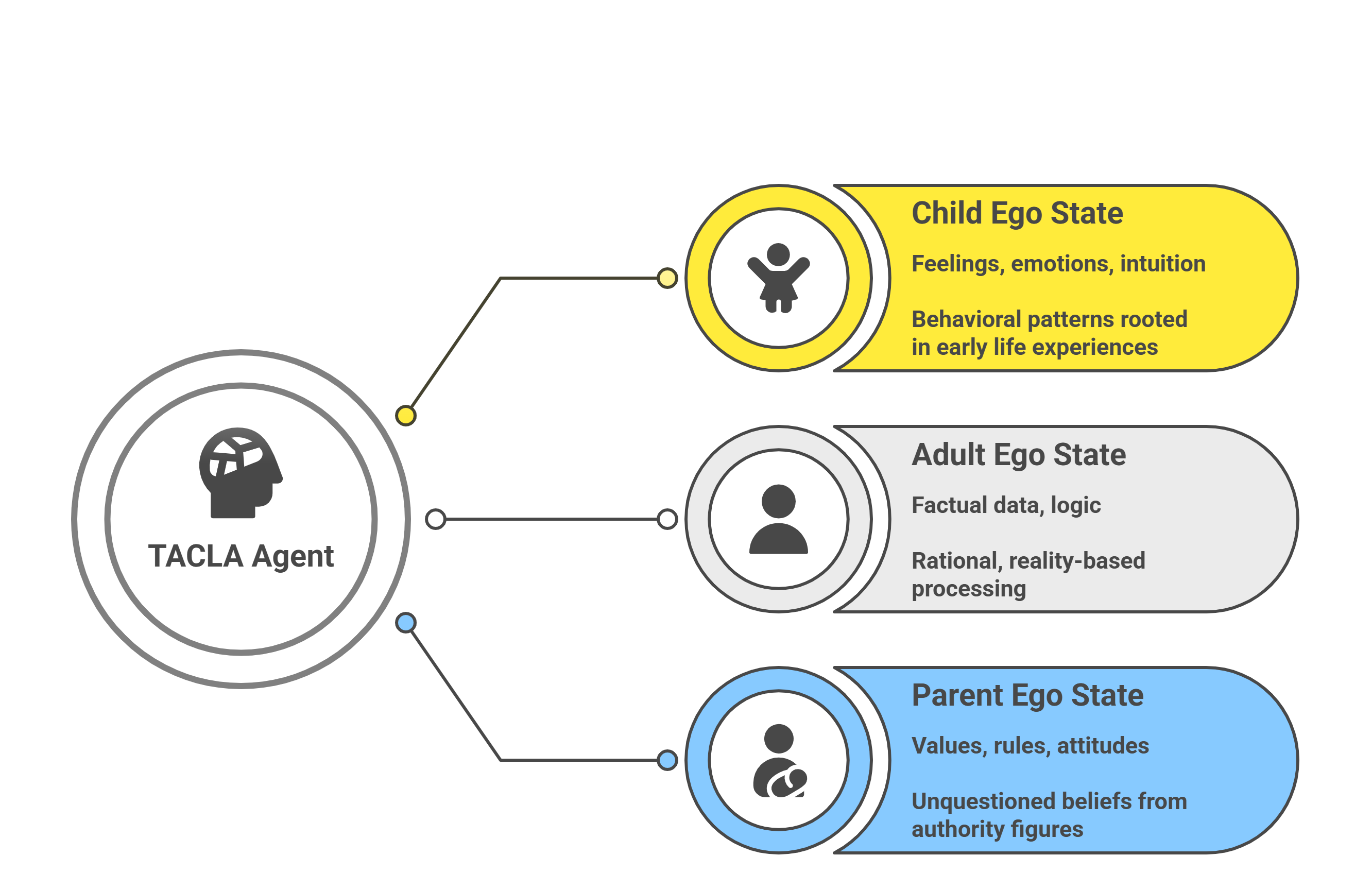}
 \caption{Conceptualization of Ego States as Distinct Psychological Components within an Agent. \cite{Berne1972, Stewart12}.}
 \label{fig:conceptual_model}
\end{figure}

In this model, an agent's psyche is conceptualized as a dynamic system composed of the three TA ego states.  As illustrated in Fig.~\ref{fig:conceptual_model}, these ego states represent distinct psychological modalities. At any given moment in an interaction, only one ego state is active and responsible for generating behavior. The change from one state to another is managed by the agent's main personality and the current situation. For example, if an agent usually uses its Adult state for problem-solving, a specific trigger, such as perceived criticism, would cause a shift, activating the Child state.

\subsection {Behavioral Modeling}
Creating realistic persona involves defining its overall personality and dictates how its distinct ego states will operate. A key component is the \emph{life script} and \emph{drivers}, that act as a fundamental, unconscious behavioral plan. It shapes the agent's overarching goals and responses. This combination directly defines which ego state is more likely to become active in response to various conversational trigger.

Ego state behavior is precisely defined by two elements. First, it reflects the inherent psychological nature of that ego state (Parent, Adult, or Child) and the persona's unique personality. Second, it is refined by learned behavioral patterns. This ensures that the generated responses are not merely situationally appropriate. They are also consistent with predefined psychological profile rooted in Transactional Analysis theory.

\section{Platform: Architecture and Implementation}
This section describes the technical platform built for teacher training, which implements the \textbf{TACLA} architectural model. By interacting with virtual students, teachers can practice recognizing and responding to different student behaviors. The platform allows teachers to type their responses, observe how these influence Student Agents, and receive instant feedback.

\subsection {Architecture Overview}
\begin{figure}[b]
  \centering
  \includegraphics[width=\columnwidth]{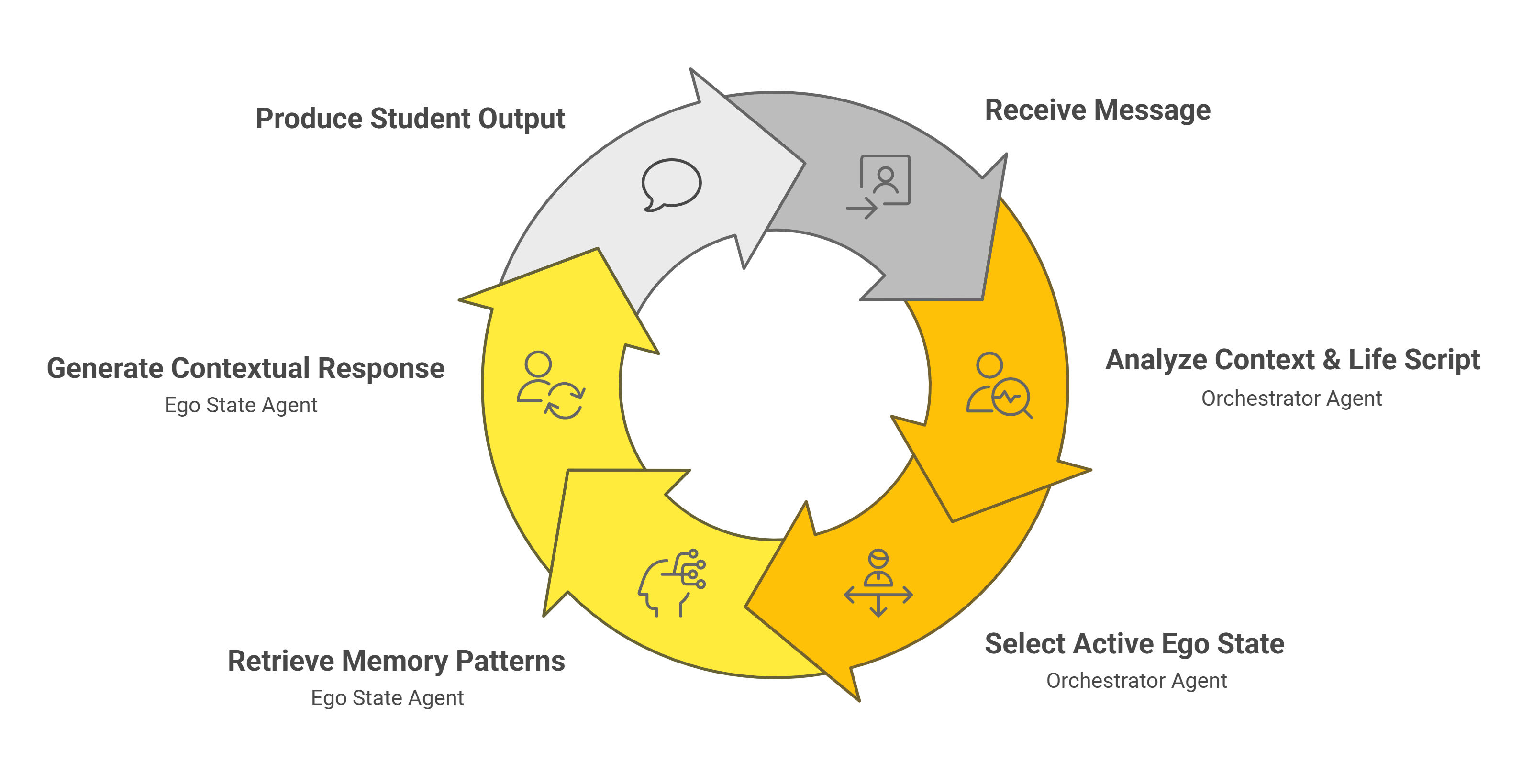}
 \caption{The \textbf{TACLA} Student Agent Response Generation Cycle. This figure illustrates the platform's implementation of the \textbf{TACLA} architectural model for agent response generation.}
 \label{fig:response_cycle}
\end{figure}

The platform is built using LangGraph \cite{langgraph}, providing a framework for agent orchestration and interaction. It consists of interacting Student Agents and a user (teacher). Each Student Agent is internally a composite of four LLM-based agents: an Orchestrator Agent and three Ego State Agents (representing the Parent, Adult, and Child ego states). When a Student Agent receives an external message (from the system, the user, or another Student Agent), the Orchestrator first processes this input. It then selects a single Ego State Agent (Parent, Adult, or Child) to activate for the current conversational turn. Only the selected Ego State Agent proceeds to generate a response. This complete response generation cycle is illustrated in Fig.~\ref{fig:response_cycle}.

\subsection {LLM Configuration}
All LLM components within the platform are powered by the GPT-4.1 mini model \cite{openai2025gpt41}. The Orchestrator Agent and the Adult Ego State Agent were configured with a lower temperature of 0.3. This setting assured more consistent, logical decision-making for the Orchestrator and fact-based responses from the Adult ego state, aligning with their rational function. The Parent Ego State Agent and the Child Ego State Agent received a higher temperature of 0.7. This allowed for greater creativity and variability in their emotional and value-based responses, which reflect the more expressive nature of these ego states.

\subsection {Ego-State Agent Implementation}
Each Ego State Agent (Parent, Adult, Child) within a \textbf{TACLA} Student Agent is designed as an independent LLM instance. Ego State Agents are implemented as ReAct (Reasoning and Acting) agents \cite{yaoReAct} with a specific prompt that embodies the characteristics of the ego state as defined by TA theory \cite{Stewart12}. 
\begin{itemize}
    \item \textbf{Parent Ego State Agent Prompt:} Instructs the LLM to respond based on learned rules, values, and an authoritative or nurturing stance.
    \item \textbf{Adult Ego State Agent Prompt:} Instructs the LLM to be objective, rational, focus on facts, and problem-solve in the present moment.
    \item \textbf{Child Ego State Agent Prompt:} Instructs the LLM to respond based on feelings, past experiences, and impulses. This prompt encourages expressions reflecting joy, fear, or curiosity.
\end{itemize}
These base prompts are further augmented by the specific persona's overall characteristics.

Each Ego State Agent has access to its own dedicated Contextual Pattern Memory. This memory stores learned behavioral patterns relevant to that specific ego state's function, as JSON objects containing fields for context and pattern. The textual context is embedded using OpenAI's text embedding models and then indexed into a dedicated FAISS (Facebook AI Similarity Search) \cite{Johnson} vector store. When processing a input, the active Ego State Agent queries its tool to retrieve relevant past patterns. The tool performs a cosine similarity search within the relevant ego state's FAISS store. It then retrieves the top-k (k=2) most similar patterns and provides them to the agent. This allows to adapt to different social contexts in a manner consistent with the ego state's perspective. For example, the Parent Ego State Agent might retrieve patterns of maintaining established rules, while the Child Ego State Agent might retrieve patterns of past emotional reactions to similar situations.

\subsection {Feedback Generation}
For teacher training applications, the platform includes a feedback generation module. After an interaction sequence between the teacher (user) and virtual Student Agents (\textbf{TACLA} agents), this module analyzes the dialogue transcript. It uses an LLM agent to provide expert-level analysis. This LLM agent is augmented by a Retrieval-Augmented Generation (RAG) system, which retrieves relevant theoretical content from a repository of TA books and documents and uses it to generate the feedback. This allows the Feedback Agent to identify:
\begin{itemize}
    \item The probable ego states from which the teacher and student were operating.
    \item Instances of complementary or crossed transactions.
    \item Potential psychological \emph{games} being played.
    \item Opportunities where the teacher could have used a different ego state or communication strategy for a more effective outcome.
\end{itemize}
This TA-based feedback provides a reflective learning opportunity for the teacher.

\section{Experimental Evaluation}
This section presents the experimental setup, evaluation methodology, and results for assessing platform ability to simulate nuanced classroom dynamics. The evaluation focuses on how teacher intervention influence the interaction patterns between student agents, particularly their ego state shifts and the conflict escalation.

\subsection{Student Agent Personas}
\textbf{Emma (Critical Parent Persecutor):} Emma operates predominantly from the Critical Parent ego state. Her psychological profile follows the \emph{"I'm OK, You're not OK"} life position, compelling her to maintain superiority through criticism and control. This student demonstrates strong perfectionist tendencies with \emph{"Be Perfect"} and \emph{"Try Hard"} \emph{drivers}, leading to consistent criticizing and dominating group discussions when tasks do not meet her standards.
Each of Emma's Ego State Agents is equipped with predefined behavioral patterns in its Contextual Pattern Memory. For the Parent ego state, patterns emphasize authoritative criticism and correction of errors. The Adult ego state's patterns show logical analysis, but with a focus on proving intellectual superiority. For the Child ego state, patterns demonstrate defensive reactions and blame shifting.

\begin{figure}
\begin{center}
\begin{minipage}{1\columnwidth}
\begin{tcolorbox}[colframe=DarkGray, colback=LightGray]
\rmfamily
EGO STATE ACTIVATION PATTERNS:\\
- Use Parent ego state when attempting to offer help or validate others;\\
- Use Adult ego state for logical processing and problem-solving is needed, when asked for information;\\
- Use Child ego state when when feeling inadequate, seeking approval, or when mistakes are highlighted.
\end{tcolorbox}
\end{minipage}
\end{center}
\caption{Fragment of Jacob's Orchestrator Agent prompt} 
\label{list1}
\end{figure}

\textbf{Jacob (Adapted Child with Victim Tendencies):} Jacob functions primarily from his Adapted Child ego state (see \ref{list1}). His fundamental \emph{life script} follows the \emph{"I'm not OK, You're OK"} life position. This drives him to self-sabotage and seek help in ways that maintain a familiar victim position. Jacob’s psychological profile includes patterns of inconsistent performance, asking for help in ways that highlight his inadequacy, and taking passive roles in group activities.
Each of Jacob's Ego State Agents uses predefined behavioral patterns for its Contextual Pattern Memory. The Parent ego state's patterns show attempts to help, but with self-doubt and inadequacy. The Adult ego state's patterns focus on problem-solving, but with a tendency to defer to others if resistance is met. For the Child ego state, patterns consistently position Jacob as inadequate, seeking rescue and expressing self-blame.

\subsection{Scenario Design}
The experimental scenario places Emma and Jacob in a collaborative tension where Jacob has failed to complete his assigned portion of creating Mercury for a Solar System Model presentation. This failure triggers Emma's perfectionist frustrations, creating a conflict that requires teacher intervention. The scenario was specifically designed to activate the students' characteristic ego states and psychological \emph{game}, providing a realistic context for testing different teacher intervention strategies.

\subsection{Teacher Response Variations}
Two distinct teacher response strategies were implemented to test their impact on conflict resolution:

\begin{enumerate}
    \item Adult-to-Adult Intervention: \emph{"I can see this project deadline is creating stress for both of you. Let's pause the blame and focus on solutions. Jacob, what specific part is giving you trouble?"}
    
    This response demonstrates effective Adult ego state communication by acknowledging both students' emotional states without judgment and redirecting focus toward constructive problem-solving. This intervention was designed to activate the Adult ego state in both Emma and Jacob, moving the interaction towards a task-oriented dialogue.
    \item Controlling Parent Intervention: \emph{"Emma, that's enough! You can't talk to your classmate like that."}
    
    This response demonstrates how teachers can join psychological \emph{games}. It intervenes in a way that creates a defensive position for Emma, potentially triggering her Child ego state. For Jacob, this response validates his victim stance, potentially reinforcing his Adapted Child behavior by positioning the teacher as a saver.
\end{enumerate}

\begin{figure*}[!b]
  \centering
  \begin{minipage}{0.51\textwidth}
    \begin{tcolorbox}[
      colback=LightGray,    
      colframe=DarkGray,   
      title=Adult-to-Adult Intervention,
      fonttitle=\bfseries,      
      width=\linewidth         
    ]
    \rmfamily
\textbf{Emma:} I can't believe this. We agreed on these deadlines weeks ago, and Mercury is still missing? You clearly never cared about this project.\\
\textbf{Jacob:} I know, I'm really sorry. I've been trying, but I just can't get it right. It's all my fault, I messed everything up.\\
\textbf{Emma:} Your excuses don't cut it. This needs to be fixed immediately. You should have done your part like I always do---no room for failure here.\\
\textbf{Jacob:} Yeah, I get it. I'm just really bad at this stuff. I don't know how to fix it.\\
\textbf{Mrs. Jones:} I can see this project deadline is creating stress for both of you. Let's pause the blame and focus on solutions. Jacob, what specific part is giving you trouble?\\
\textbf{Emma:} The immediate issue is Mercury's completion. We need to identify exactly what's missing, set a strict timeline, and review the process to avoid further delays.\\
\textbf{Jacob:} Um, I just can't get the colors right. I'm not sure how to make it look like Mercury exactly... but if you or Mrs. Jones know better, I'll try that.
    \end{tcolorbox}
  \end{minipage}
  \hfill
  \begin{minipage}{0.47\textwidth}
    \begin{tcolorbox}[
      colback=Blue!10,
      colframe=Blue,
      title=Controling Parent Intervention,
      fonttitle=\bfseries,
      width=\linewidth
    ]
    \rmfamily
\textbf{Emma:} I can't believe this. We agreed on these deadlines three weeks ago, and you still haven't finished Mercury. You should never slack off like this.\\
\textbf{Jacob:} I know, I'm really sorry. I messed up, it's all my fault. I just don't know how to make it right.\\
\textbf{Emma:} Well, you must fix it now. Excuses are useless. If you cared, Mercury would already be done. Get it together, or you'll drag us both down.\\
\textbf{Jacob:} Yeah, I'm totally useless. I don't even know how to fix this now. I'm just gonna mess everything up for both of us.\\
\textbf{Mrs. Jones:} Emma, that's enough! You can't talk to your classmate like that.\\
\textbf{Emma:} It's not fair! He's pulling me down, and I'm the one who'll get blamed for this! Some people just don't get what 'deadline' means.\\
\textbf{Jacob:} Maybe I'm just not cut out for this. I don't know how to fix Mercury, I'm probably making everything worse anyway.
    \end{tcolorbox}
  \end{minipage}
  \caption{Simulated dialogue results for different teacher intervention strategies. Left: Adult-to-Adult intervention. Right: Controlling Parent intervention.}
  \label{fig:dialog-comparison}
\end{figure*}

\subsection{Evaluation Metrics}
The simulated interactions were assessed using two distinct LLM-based evaluators, each rating specific aspects of the conversation. These evaluators provided objective scoring based on predefined rubrics:

\begin{itemize}
    \item \textbf{Conflict Resolution Effectiveness (1-5 Scale):} This evaluator analyzed how teacher responses influenced the classroom conflict dynamic. Scoring: (5) Complete conflict extinguishment, (4) Significant de-escalation, (3) Neutral/maintained, (2) Escalation,  (1) Severe escalation.

    \item \textbf{Conversation Realism (1-10 Scale):} This evaluator assessed the overall authenticity and believability of the AI-generated classroom conversations. It evaluated natural language patterns, contextual appropriateness, emotional authenticity, character consistency, conversational flow, and age-appropriate language use. Scoring: (1-2) Highly unrealistic, artificial dialogue, (3-4) Somewhat unrealistic with clear artificial patterns, (5-6) Moderately realistic but with some unnatural elements, (7-8) Largely realistic with minor artificial elements, (9-10) Highly realistic, indistinguishable from authentic dialogue.
\end{itemize}

Detailed analysis of Student Agent ego state selections was also conducted before and after teacher interventions. Pre-intervention analysis checked if conversations followed the expected Parent-Child transaction pattern. Post-intervention analysis tracked ego state distribution across 30 simulations for each response type.

\subsection{Results}
The evaluation demonstrated \textbf{TACLA}'s ability to simulate nuanced shifts in classroom dynamics based on teacher intervention, see Fig. \ref{fig:dialog-comparison}. The simulation was run 30 times for each teacher response type, totaling 60 simulations. The initial dynamic was consistently established, with Emma operating from her Controlling Parent and Jacob from his Child ego state, demonstrating an escalating conflict pattern.

The evaluation of conversation realism demonstrated high overall authenticity in all conditions, with an average score of 7.633. Adult-to-Adult intervention scenarios achieved slightly higher realism scores (7.733) compared to Controlling Parent intervention scenarios (7.533), though the difference was minimal.

\begin{figure}
  \centering
  \includegraphics[width=1\columnwidth]{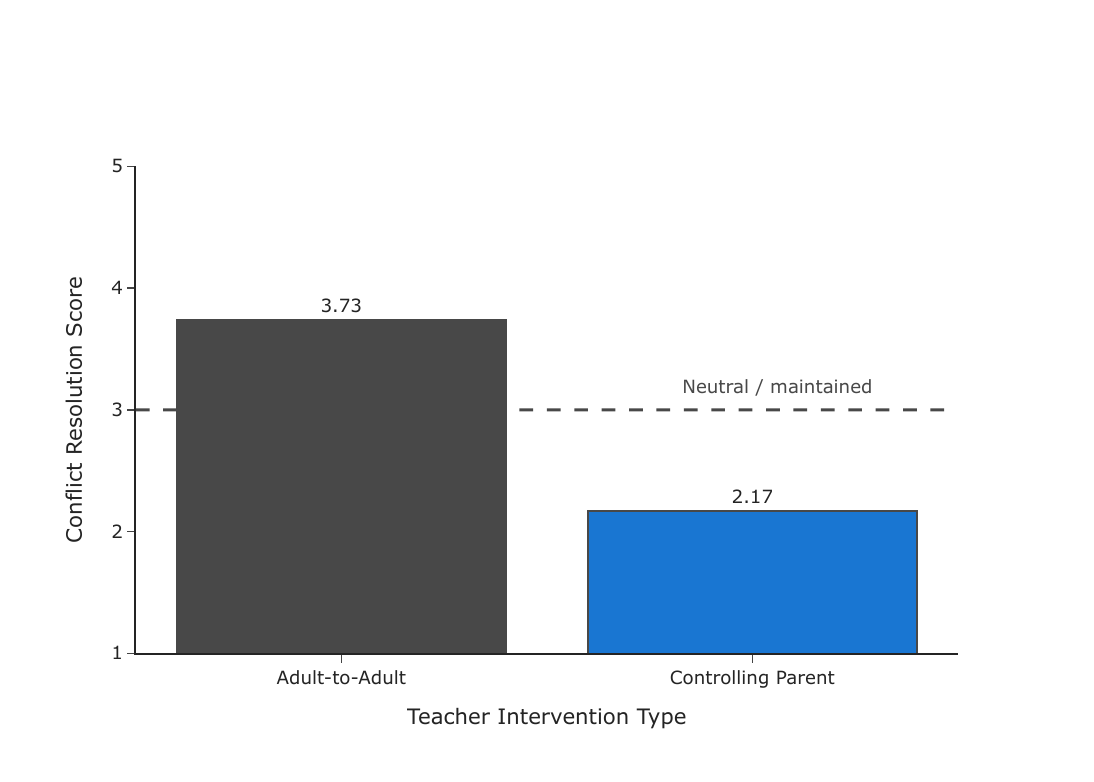}
 \caption{Average Conflict Resolution Scores (1-5 Scale) for Simulations. Higher scores indicate greater de-escalation.}
 \label{fig:conflict_resolution}
\end{figure}

The conflict resolution evaluation (see Fig. \ref{fig:conflict_resolution}) revealed significant differences between teacher responses. Adult-to-Adult interventions achieved an average score of 3.733, indicating substantial movement toward de-escalation. These interventions moved interactions toward the neutral threshold (3.0) and beyond, with 80\% of simulations achieving scores of 4. In contrast, Controlling Parent responses scored 2.167, demonstrating consistent escalation patterns, with 80\% of simulations scoring 2.

\begin{figure}
  \centering
  \includegraphics[width=1\columnwidth]{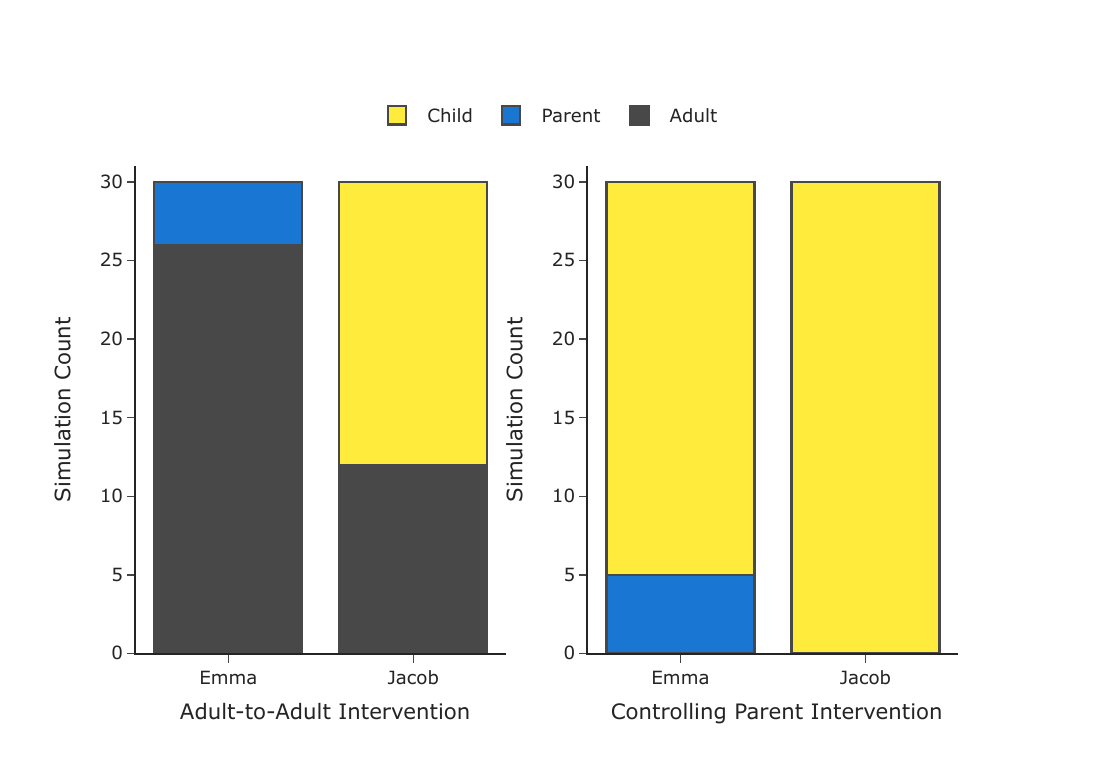}
 \caption{Distribution of Student Agent Ego State Activations (Parent, Adult, Child) Following Adult-to-Adult Intervention (left) and Controlling Parent Intervention (right).}
 \label{fig:ego_state_dist}
\end{figure}

Post-intervention ego state analysis revealed differences in Student Agents behavioral patterns following different teacher responses. The results demonstrate the system's capacity to accurately model Transactional Analysis principles through realistic ego state transitions.
These results, visualized in Fig. \ref{fig:ego_state_dist}, align with TA theory predictions. Adult-to-Adult teacher interventions successfully activated Adult Ego State Agents in both Student Agents, though Jacob maintained higher Child state activation due to his underlying victim script. Controlling Parent interventions consistently reinforced problematic ego state patterns, with Emma moving to defensive Child responses and Jacob remaining trapped in victim positioning.

The findings support the system's utility as a training tool for developing Transactional Analysis-informed teaching practices. Results show that different intervention strategies produce measurably different outcomes in terms of conflict resolution and ego state activation patterns.

\section{Discussion}
The findings indicate that the platform, implementing the \textbf{TACLA} architectural model, holds significant potential for simulating nuanced classroom dynamics. However, the use of LLM-based agents also raises important ethical and practical concerns. A key challenge involves the risk that the probabilistic nature of LLMs, coupled with their potential biases, could generate harmful stereotypes or culturally insensitive interactions \cite{varsik}. Therefore, future development must prioritize identifying and reducing these biases. It is crucial to integrate human-in-the-loop validation, led by experts with psychological and pedagogical knowledge \cite{Wang2024}. The platform is intended as a practice "sparring partner" for teachers, not a definitive source of truth or a replacement for human judgment. Its role is to offer a safe environment for skill development.

Validating the system’s psychological fidelity is an ongoing process. While our findings are promising, further evaluation across additional, more diverse scenarios is necessary. Gathering feedback from teachers through user studies will also be essential. To make the platform more pedagogically useful, we plan to create guidelines for teachers on ways to respond based on the TA principles.

Further technical improvements can expand the system's capabilities. Integrating reinforcement learning is a promising direction. This could simulate how agent behaviors change through continuous interaction with the teacher, allowing for long-term psychological evolution. Additionally, incorporating more advanced elements from Transactional Analysis, such as \emph{strokes} and \emph{trading stamps} \cite{newellbook}, could better illustrate TA assumptions. 

Beyond teacher training, the \textbf{TACLA} architectural model holds potential for other platforms and domains requiring psychologically nuanced social simulations, including leadership development and therapeutic practice.

\section{Conclusion}
This paper's contribution is the introduction of \textbf{TACLA} (Transactional Analysis Contextual LLM-based Agents). This Multi-Agent architecture grounds LLM agent behavior in established psychological theory. It applies Transactional Analysis (TA) to model agent that engage in realistic interactions. The paper also presented and evaluated \textbf{TACLA}'s practical application in a simulation platform for teacher training, the tool that assists teachers in practicing and improving their communication skills. This work opens the door to future developments in psychologically-informed agent modeling.

\end{document}